\documentclass[twocolumn,rmp,aps]{revtex4}

\usepackage{psfig,graphics,graphicx}

\begin{document}

\title{Experiments in vortex avalanches}

\author{E. Altshuler}
\email{jea@infomed.sld.cu}
\affiliation{Superconductivity Laboratory 
and ``Henri Poincar{\'e}" Group of Complex Systems,
IMRE-Physics Faculty, University of Havana, 10400 Havana, Cuba}
\author{T. H. Johansen}
\email{t.h.johansen@fys.uio.no}
\affiliation{Department of Physics, University of Oslo, Blindern,
N-0316 Oslo, Norway}

\begin{abstract}

Avalanche dynamics is found in many phenomena spanning from
earthquakes to the evolution of species. It can be also found in
vortex matter when a type II superconductor is externally driven,
for example, by increasing the magnetic field. Vortex avalanches
associated with thermal instabilities can be an undesirable effect
for applications, but ``dynamically driven" avalanches emerging
from the competition between intervortex interactions and quenched
disorder constitute an interesting scenario to test theoretical
ideas related with non-equilibrium dynamics. However, 
differently from the equilibrium
phases of vortex matter in type II superconductors, the study of
the corresponding dynamical phases --in which avalanches can play
a role-- is still in its infancy. In this paper we critically
review relevant experiments performed in the last decade or so,
emphasizing the ability of different experimental techniques to
establish the nature and statistical properties of the observed
avalanche behavior.

\end{abstract}

\date{\today}
\maketitle

\tableofcontents

\section{INTRODUCTION}
\label{sec:intro}

Rooted somewhere between Physics and Engineering, the critical
state model of Charles P. Bean (1962) continues to enjoy an immense
popularity amongst those who need to understand the magnetic
properties of almost all potentially useful superconductors.
Above a certain magnetic field threshold, type II superconductors
are penetrated by superconducting {\it vortices}, or flux lines,
each one consisting in a normal-state core surrounded by a tiny
supercurrent tornado with a few-dozen-nanometer radius. The
vortices can therefore be thought of as long and thin solenoid
magnets which enter into the sample in increasing numbers as the
external field grows. In a perfect superconducting crystal, the
competition between the inter-vortex repulsion and the ``magnetic
pressure" from the outside field causes the vortices to arrange
in a hexagonal lattice \cite{Abrikosov:1957}. In a real
superconductor, however, there are defects acting as pinning
centers, and the vortex motion becomes impeded. The interplay
of all these forces, where an external drive ``pushing in" more and
more vortices is counteracted by pinning, results in a
non-equilibrium state, the {\it critical state}, with a vortex
density being largest near the surfaces where flux enters the
sample. This critical state typically involves several million
vortices, and as the external field is increased or decreased,
they readily find the way to organize themselves in spite of
their short range interactions.
Researchers in the area of Complexity would not hesitate 
these days in characterizing Bean's critical state 
as an {\it emergent
phenomenon} resulting from the {\it self organization} of a
{\it complex system} of vortices.

But those are not empty words. They call attention to the fact
that the collective, nonlinear statistical properties of a complex
system can produce amazing macroscopic results, regardless of the
details of the interaction between their microscopic constituents.
They also suggest that we should open up our mind and try to find
analogies in eventually very distant fields of science. After all,
isn't a sandpile a quite good analogue of Bean's critical state?
As grains are added to a sandpile from the top, gravity tries to
bring them off the pile, a motion which is prevented by the
intergrain friction. And again, in spite of the short range
character of the latter, the pile finds the way to organize itself
and produce globally an {\it angle of repose}, or critical
angle. In very simple terms, you can identify the gravity as the
magnetic field applied to superconductors, while friction
corresponds to vortex pinning. This would have sounded like
celestial music in the ears of Lord Kelvin, who once wrote ``I am
never content until I have constructed a mechanical model of the
subject I am studying. If I succeed in making one, I understand;
otherwise I do not." \cite{Kelvin:1884}.

Grasping the analogy between the critical state and the sandpile
well before the era of Complexity ideas, Pierre G. DeGennes expresses
in his classic 1966 book {\it Superconductivity of Metals and Alloys}:
``We can get some physical feeling of this critical state by thinking
of a sand hill. If the slope of the sand hill exceeds some critical value,
the sand starts flowing downwards (avalanche). The analogy is, in fact,
rather good since it has been shown (by careful experiments with pickup
coils) that, when the system becomes over-critical, the lines do not move
as single units, but rather in the form of avalanches including typically
50 lines or more" \cite{DeGennes:1966}. This picture was dormant for many
years until scientists working in the field of Complexity identified
{\it avalanche dynamics} as a major mechanism in many  physical, chemical,
biological and social phenomena. In particular, the ideas of Self Organized
Criticality (SOC) find avalanches with ``robust" power-law distributions
of sizes and durations, at the core of the underlying dynamics in many
systems \cite{Bak:1996,Jensen:1998}. With the sandpile being a central
paradigm of SOC theory, Bean's critical state has become a natural
playground
to look for avalanche dynamics. Although heroic efforts were made in the
1960s to see these avalanches, it was computer-controlled data acquisition
that made it possible to investigate vortex avalanche {\it statistics} in
superconductors. Other advances such as micro Hall probes and high
resolution magneto-optical imaging have finally stamped the seal of
contemporary times on these studies. ``Dynamically driven" avalanches like
the ones suggested by the sandpile analogy can, after all, be the intrinsic
mechanism in the formation of the critical state that Charles P. Bean
would have never dreamed of.

In Bean's time, however, another kind of vortex avalanche attracted
most of the attention: {\it flux jumps}. Instead of helping to establish
the critical state, they tend to destroy it, producing undesirable jumps
in the sample magnetization. In contrast to the avalanches discussed in
connection with sandpiles, flux jumps are {\it thermally triggered}. If the
external field is increased too fast, and the thermal capacity and
conductivity of the sample are small, the vortices rushing in will
dissipate heat due to their motion, and the local temperature rises. This
tends to detach other vortices from their pinning sites, leading to new
motion that can cause even further heating. This positive feedback process
may sweep away the critical state in a big region of the sample, and
translates into a sudden, catastrophic decrease in the magnetization.
Thermally triggered avalanches have long been modeled in terms of
macroscopic parameters. However, present imaging techniques have provided
data showing that these events sometimes also result in complex magnetic
spatial structures which deserve a more detailed explanation.

All of these findings suggest that the simple sandpile analogy of
vortex avalanches must be examined with caution: For one thing,
temperature is not accounted for in the standard SOC picture. At
this point, many questions arise: Can experiments reveal a sharp
difference between dynamically and thermally driven avalanches? If
so, can we through statistical analyses of the dynamically driven
avalanches conclude whether Bean's critical state model represents
an SOC phenomenon? What is the relation between the details of the
magnetic flux distribution inside the sample and the avalanche
dynamics? Some authors have directly aimed their experimental
efforts at these subjects. Others offer relevant data just as
experimental ``side effects." The fairly few available outputs can
be characterized as diverse and entangled, and it is the purpose
of this Colloquium is to provide a coherent overview that highlights the
essence of the results obtained in this area during the last
decade or so.

\section{THE NATURE OF VORTEX AVALANCHES}
\label{sec:kin}

\subsection{The critical state}
\label{sec:formal}

When the external magnetic field exceeds the so-called lower
critical field, $H_{c1}$, the surface layer of a type-II
superconductor starts to give birth to vortices, which immediately
are pushed deeper into the material by the Meissner shielding
currents. Each flux line consists of a ``normal" core of radius
$\xi$, the coherence length, surrounded by a circulating
supercurrent decaying over a distance $\lambda$, the London
penetration depth. The current is accompanied by an axial
magnetic field
decaying over the same $\lambda$, and integrates to a total amount
of flux equal to the flux quantum $\Phi_0 = h/2e 
\approx 2 \times 10^{-15}$ Tm$^2$, where $h$ is Planck's constant 
and $e$ is the elementary charge. 
As the applied field keeps increasing, the vortices
get closer and closer until the overlapping is so heavy, that an
overall transition to the normal state takes place at the upper
critical field, $H_{c2}$. When microscopic defects are present in
the material, such areas tend to pin any vortex that passes by.
The pinning force always acts against the driving force, which on
a vortex has a Lorentz-like form, ${\bf f}_L = {\bf J} \times
\Phi_0 {\bf \hat{z}}$, where ${\bf J}$ is the local density of
either a transport current or a magnetization current (Meissner
current and/or gradients in the number of neighboring vortices),
or both.
The basic assumption of the critical state model is that as the
vortices invade the sample, every pinning center that catches a
vortex will hold onto it as hard as it possibly can, quantified
by a certain maximum pinning force per unit vortex length,
$f_p^{\rm max}$. In this way the local balance between the two
competing forces, $|{\bf f}_L| = f_p^{\rm max}$,  creates a
metastable equilibrium state, where the current density adjusts
itself to a maximum magnitude, $|{\bf J}^{\rm max}| \equiv J_c$,
the critical current density.
>From Ampere's law it then follows that the flux density distribution,
${\bf B(r)}$, in the critical state is given by
\begin{equation}
\mid \nabla \times {\bf B(r)}\mid = \mu_0 J_c .
\label{eq:Bean1}
\end{equation}
The vortices therefore organize in such a way that their density
decreases linearly from the edges of the sample, and the slope is
$\mu_0 J_c$, as illustrated in Fig.~1(a). Shown in Fig.~1(b)
is a set of $B$-profiles that occur at different stages during an
ascent (left) and the subsequent descent (right) of the applied
field. From the illustration it is evident that this strongly
hysteretic process is quite analogous to what happens to a box of
sand where sand is first added near the side walls (left), and
afterwards the walls are gradually lowered to zero height (right).
The crux is then: How do such systems evolve in space and time
as they are driven externally through a  ''continuous sequence''
of different critical states?

\begin{figure}
\centerline{\psfig{figure=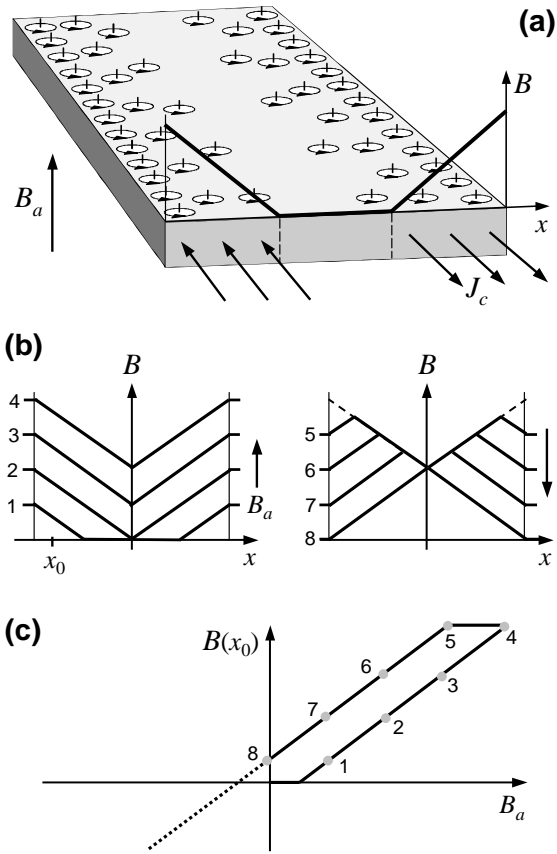,width=8cm}}
\caption{Bean's critical state; (a) the distribution of
vortices, internal field and current in a superconductor
placed in an external magnetic field, $B_a$.
(b) internal field profiles for increasing (left) and
decreasing (right) $B_a$, and (c) variation of the local
field at $x_0$ during the cycle in (b).
\label{f_f1}}
\end{figure}

\subsection{Dynamically and thermally driven avalanches}

The idea of dynamically driven avalanches in the vortex matter
relates to one possible way for the system to respond
when subjected to a {\it slow} drive,
e.g., a gentle ramping of the applied magnetic field.
By driving the vortices sufficiently slowly only their mutual
repulsion and the interactions with pinning sites
are expected to control the dynamics.
If SOC provides the correct description, the critical
state behavior should show scale invariant avalanche dynamics,
i.e., the distribution of avalanche sizes follows a power law,
$P(s) \sim s^{-\alpha}$. Here $P(s)$ is the probability
to find an avalanche event where $s$ vortices suddenly move,
and $\alpha$ is a critical exponent. While in 
the original formulation 
of SOC the exponent $\alpha \approx 1$ is found to be robust  
with respect to small changes in the model, later developments 
of the theory have shown that the exponent can vary within a 
certain range. 

Note that in some cases the finding of temporal signals 
exhibiting scaling, e.g., signals with $1/f$ noise in 
the power spectrum, has been taken as 
direct evidence for SOC behavior.
However, observation of $1/f$ noise should not be considered a sufficient 
indication of SOC, 
since it can result even from a 
 spread of activation 
energies \cite{Obrien:1992,Jensen:1998}.

Whether real piles follow the SOC scheme is still
subject to debate \cite{Held:1990,Bretz:1992,Rosendahl:1993,
Rosendahl:1994,Frette:1996,Altshuler:1999}, and a similar 
discussion extends also to several other systems 
\cite{Field:1995,Plourde:1993,Danna:2000}.
It is therefore
important to note that the critical state of type II
superconductors represents a unique and attractive case to study.
In contrast to grains of sand, the vortices are non-inertial
objects, and are hence closer to the idealized formulation of the
SOC theory.

As in most areas where SOC ideas have been applied, the
theoretical papers largely outnumber the experimental studies of
vortex avalanches. Let us therefore, as a background for the main
part of this Colloquium, mention briefly the important trends in the
theoretical work, emphasizing ideas and results that most directly
connect to the available experiments.
Among computer simulation there are two philosophies 
dominating the literature; molecular dynamics (MD)
and cellular automata (CA). In addition, a few reports using a 
macroscopic approach have been published. 


Most macroscopic treatments discuss vortex
avalanches in a thermal activation scenario \cite{Tang:1993,Bonabeau:1995,
Bonabeau:1996,Pan:1994,Prozorov:1999,Vinokur:1991}. 
Although some of these authors claim to find fingerprints of SOC behavior, 
their results are not compatible with the ``canonical" formulation by 
Bak {\it et al.} (1987): As in a shaking sandpile, thermal activation
makes the critical state to relax {\it away} from the marginal stability
because vortices, or bundles of them, jump out of their pinning centers, 
and redistribute in such a way that the Bean's profile changes in time.
This phenomenon, known as {\it flux creep}, was first observed
by Kim {\it et al.} in 1963, and its typical manifestation is a {\it slow},
logarithmic temporal decay of the magnetization
(Yeshurun, 1996). Thus, flux creep
can only be allowed within a ``soft" definition of SOC, eventually
useful to interpret certain relaxation experiments which will be
discussed later in this Colloquium (\cite{Aegerter:1998}).
There are also macroscopic studies that do ignore flux creep effects
\cite{Barford:1997}. Here, the author proposes an
equation of motion to analyze the dynamics of the critical state
as the external field is increased, and finds a
power law in the distribution of avalanche sizes with
a critical exponent of $1.13$, consistent with the original SOC picture. 

Typical for the MD simulations is that they allow integration of the
equations of motion at the vortex level. Since this demands quite high
computing power, the MD work deals mostly with small systems.
The CA approach, on the other hand, simplifies the dynamics by
selecting a set of physically sound rules that imitate the
real laws, thereby allowing simulation of much bigger systems.
Care must be taken, however, since the results can be sensitive
to the selected set of rules (see, for example, Kadanoff {\it et al.}(1989)).

After the pioneering application of MD techniques in the investigation
of vortex avalanches in the critical state by Richardson {\it et al.} (1994),
extensive work on the subject has been 
generated \cite{Barford:1993,Olson:1997,Olson:1997a,Pla:1996}. A MD simulation
of a slowly driven critical state can be illustrated
by the approach of Olson {\it et al.} (1997): For every vortex,
{\it i}, they solve the overdamped equation of motion
\begin{equation}
{\bf f}_i={\bf f}_{i}^{vv}+ {\bf f}_{i}^{vp}=\eta {\bf v}_i,
\label{MD}
\end{equation}
where ${\bf f}_i$ is the total force, comprised of ${\bf
f}_{i}^{vv}$ -- the intervortex repulsion, and ${\bf f}_{i}^{vp}$
--  the interaction between the vortex and a pinning center. The
${\bf v}_i$ is the vortex velocity, and $\eta$ the ``viscosity" of
vortex flow. With this realistic description of each member of the
ensemble the simulations show that a critical state flux profile
builds up when vortices are slowly added
from one side of the ``sample". If one keeps adding vortices 
after the critical state
is fully established, their effect can be followed by calculating
the time evolution of the average vortex velocity. Typically, this
shows {\it bursts} of activity, or avalanches, which resemble the
voltage signals found in the pick up experiments discussed below
\cite{Field:1995}. The avalanche size distribution resulting from
these simulations follows a power law. When counting all the 
moving vortices for each
avalanche event, one finds an exponent in the range
$0.9 \le \alpha \le 1.4$, where the spread comes from varying the
strength and density of the pinning sites. Similarly, for
off-the-edge avalanches (counting only the number of vortices
exiting through the ``sample edge" during an event) one finds $2.4
\le \alpha \le 4.4$. Although these distributions are often
well-behaved over quite a broad range, it is also clear that the
exponent is not very robust. 


\begin{figure}
\centerline{\psfig{figure=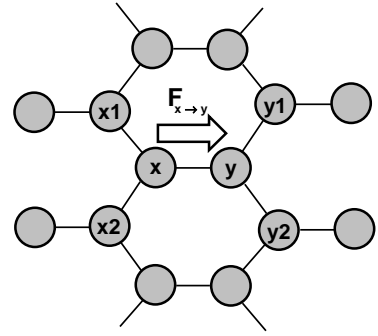,width=5.5cm}}
\caption{Site lattice illustrating the Bassler-Paczuski cellular
automata used for modelling vortex avalanches \cite{Bassler:1998}.
\label{f_f2}}
\end{figure}

The CA approach was introduced in a model by Bassler and Paczuski
(1998), where they considered vortex dynamics on a two-dimensional
honeycomb lattice (see Fig.~2). Each cell $x$, which has three nearest
neighbors, is occupied by an integer number of vortices. 
The authors then assume that force pushing a vortex at $x$ towards the 
neighbor cell $y$ consists of two basic contributions:
Firstly, to mimic the vortex-vortex repulsion the force increases as
the population at $x$ gets bigger than that at $y$. A similar term 
representing the next nearest neighbor repulsion is also included.
Secondly, to simulate the vortex-pin
attraction, the force increases as the pinning potential
of $y$ is bigger $x$ (the pinning potential is represented by 
a random number 
assigned to each cell). 
In each time step, the
cells are updated in parallel; a vortex moves to a neighboring
cell if the force in that direction is positive. If a vortex is
attracted in more than one direction, the selection can be made at
random \cite{Bassler:1998} or by a largest force rule
\cite{Bassler:2001}. When vortices are now added at one edge of
the ``sample," this CA leads to a critical state very close to the
ideal Bean's flux profile. As in the MD simulations, one finds
here avalanche dynamics, and the size distribution of avalanches
was reported to have a critical exponent of $1.63 \pm 0.02$
(obtained after finite size scaling for four orders of magnitude
in avalanche size). In contrast to the MD work, the exponent is
here essentially constant within the range of parameters studied, 
therefore suggesting a SOC scenario
\cite{Bassler:1998}. The application of the model to the case of
periodic, dense pinning, indicates a slight decrease in the
exponent to $1.45 \pm 0.02$ \cite{Cruz:1999}.

Besides SOC there are also other theories producing power laws in 
the avalanche size distributions 
\cite{Carlson:1999,Schwarz:2001,Newman:1996,Huang:1997}.
Among these, the model of Newman and Sneppen (1996) seems the most 
relevant to the critical state, although 
the excitation in the form of ``coherent noise" is not 
obviously applicable to vortex dynamics.


Catastrophic avalanches --flux jumps-- are
associated with a ``runaway" in the motion of vortices as they
redistribute in response to e.g. an increasing applied field. Per
unit volume, the motion generates heat at the rate of $J_cE$, where
$E$ is the electrical field. Due to this dissipation, the
critical current density and thereby the shielding goes down, and
more vortices rush into the sample. This positive feedback may or
may not result in a flux jump. The superconductor is stable if the
heat dissipation does not exceed the material's ability to store
heat, a criterion that under adiabatic conditions can be expressed
as \cite{Mints:1981}
\begin{equation}
\label{JumpsBeta} \frac{\mu _{0} J_{c}(T)w^2}{c}
\; |\frac{{\rm d}J_c}{{\rm d}T}| \ \equiv \beta <1 \, ,
\end{equation}
where $c$ is the specific heat, and $w$ a
typical dimension of the sample.\footnote{A prefactor of order
unity is omitted in the formula.}
However, if $\beta > 1$, flux jumps are to be expected, and the first jump
will occur when the field reaches the value
$B_{\rm fj} \approx \sqrt{\mu_{0} c (T_c-T)}$. Here $T_c$ is the critical 
temperature, and a linear $J_c(T)$ is assumed
as a reasonable approximation.
Let us put numbers on two
 cases that will be discussed later. For the $1.5
\times 1.5$~mm$^2$-area Nb foils used in Altshuler {\it et al.}
(2002), one gets $\beta \approx 5 \times 10^{-3}$, so flux jumps
at the temperature of 4.6~K can be discarded. For the mm-sized
YBaCuO crystals studied in the sub-Kelvin range by Seidler {\it et
al.} (1993) and Zieve {\it et al.} (1996), $\beta$ becomes close
to 3, and the situation is marginal. If flux jumps were to take
place, they would here start at $B_{\rm fj} \approx 5$ Tesla,
actually not very far from the threshold fields reported by these
authors.
However, estimates like these must be seen with caution. No real
experiment takes place under ideal adiabatic conditions, so other
factors need to be considered as well.
Generally, the ``recipe" to avoid flux jumps is to choose
samples with high thermal conductivity, make sure that their
thermal contact with the environment is good, and be gentle when
ramping the applied field.

\section{EXPERIMENTAL TECHNIQUES}

The various magnetometric techniques used to measure
vortex avalanches can be classified as {\it global} and {\it local}.
The global techniques are sensitive to either the amount of flux
passing through the surface of the sample or the volume averaged
magnetic moment, whereas the local techniques are detecting the
flux density or even the individual vortex positions in selected
regions. In this section we give a brief overview of the various
methods used in these experiments.

Pick up coil detection is the most basic global technique, and is
typically configured as a coil wound tightly around the sample.
When the external field is ramped up or down, the magnetic flux
that enters or leaves the sample will (according to Faraday's law)
induce a voltage in the coil proportional to the rate of this
``traffic" of vortices. Therefore, a steady-state flux motion
results in a constant voltage output, while the appearance of
spikes in the signal implies step-like increments, i.e., vortex
avalanche events. By integrating the voltage in time one can
determine, at least approximately, also the amount of flux
involved in such events, as done in the careful experiments of
Field {\it et al.}, described in more detail below.

Another important technique is Superconducting Quantum Interference
Device (SQUID) magnetometry (Barone and Patern{\'o}, 1982).
The basic sensor here is a closed superconducting loop interrupted
by, for instance, two Josephson junctions. A dc bias current is
injected in such a way that it flows through the two junctions in
parallel. If the loop is now subjected to a magnetic field,
this produces a shift of the superconducting phase difference through
the junctions, analogous to the phase difference between the various
optical paths in the Young's double slit experiment. As a consequence,
the maximum bias current that can be forced into the SQUID without
dissipation becomes field dependent:
$i_{m}(\Phi_{ext})=2I_{cj}\mid \cos( \pi \Phi_{ext}/\Phi_0) \mid$,
where $I_{cj}$ is the Josephson critical current of each junction,
and $\Phi_{ext}$ is the magnetic flux threading the SQUID loop.
The periodic form of $i_{m}$ implies that such a sensor can
``intrinsically detect" magnetic flux with a resolution of less than one
flux quantum. In practice, the field sensitivity of the SQUID depends
on the loop area, and on the design of flux transformers.
The areas of SQUID loops (or flux transformer pick up coils)
typically span from around 1~mm$^2$ to  0.04~mm$^2$ \cite{Lee:1995},
the latter making it possible to apply the device for local
measurements.

While sensors based on the Hall effect have long since
proved very powerful, it was the invention of the
modulation-doped semiconductor heterostructure \cite{Dingle:1978}
that gave rise to the present state-of-the-art sensors, the micro-Hall
probes. These epitaxial structures, mostly GaAl/AlGaAs, consist of
2D layers of electrons with large carrier mobilities at low
temperatures. The active area of the sensing element spans from
100~$\mu$m$^2$ to less than 1~$\mu$m$^2$.  Note then that if just
one flux quantum is present under a 100~$\mu$m$^2$ probe, the
effective field is $\approx 0.2$~Oe.
Typically, this produces a Hall output of 2~$\mu$V for a bias
current of 100~$\mu$A. Micro-Hall probes can today be manufactured
also as arrays of sensors in either linear or matrix arrangements.
A practical linear array is composed of 11 square probes of
100~$\mu$m$^2$ each, separated by 20~$\mu$m center-to-center
(Altshuler {\it et al.}, 2002).
Micro Hall probes can also be attached to a piezoelectric scanner tube
(as in a tunneling microscope) forming a {\it scanning Hall probe
microscope} (SHM)\cite{Bending:1999}. Such a device is able to
magnetically scan the sample with sub-micron spatial resolution,
and resolve the field from individual vortices. A limitation of
the method is that a standard SHM can scan only small areas,
typically $25 \times 25~\mu$m$^2$ at 77~K \cite{Oral:1996}.

\begin{figure}
\centerline{\psfig{figure=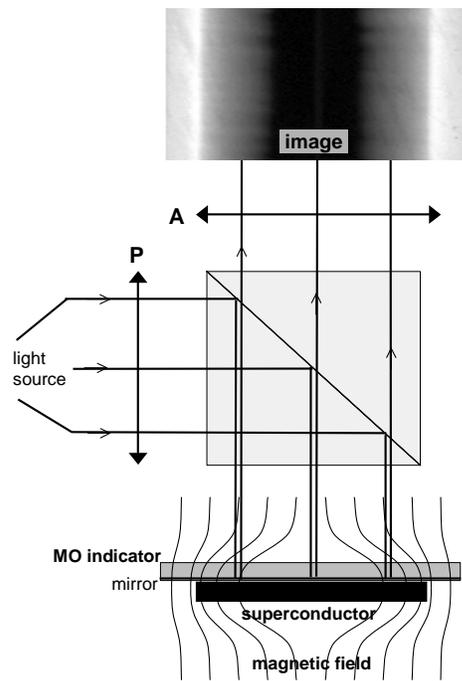,width=6.3cm}}
\caption{Principle of the magneto-optical imaging (MOI) technique. 
A magneto-optical indicator film placed
on top of the superconductor gives the incoming polarized light a
Faraday rotation according to the local magnetic field.
After being reflected and passed through a crossed analyzer, the light
produces an image in which the intensity contrast is a direct map of
the field distribution.
\label{f_f3}}
\end{figure}

The only technique which today allows experiments with combined high
spatial and temporal resolution is magneto-optical imaging (MOI).
Here the sensing element is a strongly Faraday rotating film,
which one places directly on top of the sample under investigation.
As illustrated in Fig.~3 the imaging is done by shining polarized
light through the film, where reflection from a mirror, or the
sample itself, gives the light a second pass that doubles
the Faraday effect. The light contains then a distribution of rotation
angles, $\theta_F$, corresponding to the magnetic field variations
across the face of the superconductor. Finally, an analyzer set at
90$^{\circ}$ crossing relative to the polarizer filters the light
and produces an optical image where the brightness shows directly
how the magnetic field was distributed.
Since MOI was invented in the 1950s several materials have been
used as indicator films \cite{Koblischka and Wijngaarden:1993}.
During the last decade the most popular material by far has been
the in-plane magnetization ferrite garnet films, often
(Lu,Bi)$_{3}$(Fe,Ga)$_{5}$O$_{12}$, grown as a few micron thick
epitaxial layer on gadolinium gallium garnet (transparent) substrates.
The sensitivity of these indicators is represented
by the low-field Verdet constant, $V =  \theta_F / {H d} $, where
$d$ is the film thickness.  For green light (strongly present in
Hg-lamps) one has $V \simeq 2-8$ degrees/kOe per micron, which
is sufficient to resolve individual vortices \cite{Goa:2001}.
The unique power of the MOI technique is two-fold; first, by simple
optical means one may zoom between cm- and micron-sized fields-of-view,
and second, the time response of the garnet film is extremely fast,
of the order of nanoseconds \cite{Runge:2000}.

\section{REVIEW OF RECENT EXPERIMENTS}

\subsection{Pick up coil experiments}

The first experiment on vortex avalanches inspired by the SOC
ideas was reported by Field and coworkers in 1995 \cite{Field:1995}.
An 1800 turn pickup coil was coaxially mounted on the inner surface
of a tube made from the conventional superconductor NbTi.
The tube had a 6~mm outer diameter, a wall thickness of 0.25~mm and
it was 3.4~cm long, nearly twice the length of the pickup coil.
As noted by Field {\it et al.} (1995), this geometry guarantees 
a close analogy to
(conical) sandpiles. An external magnetic field was applied along
the tube axis at various ramp speeds, and the voltage induced in
the pickup coil was amplified and recorded by a computer.
The upper section of Fig.~4 displays the time variation of the
signal over a field interval of 30~Oe centered at 7.55~kOe using
the fairly low ramp rate of 5~Oe/s.\footnote{An accepted experimental
meaning of a ramp rate being
sufficiently low in the search for SOC behavior, is that the
resulting avalanche statistics becomes insensitive to the actual
chosen rate. Typically, this occurs below 10~Oe/s.}
The authors identify two contributions to the flux penetration:
A first one, amounting to about 97 {\%} of the flux, corresponds to 
the background level, and is believed to represent the thermally 
activated ``smooth" flow of vortices. The second contribution
is the well-defined spikes, which clearly indicate
the presence of flux avalanches.

The lower panel of the figure shows the avalanche size distributions
obtained from such experiments performed at 3 different fields.
In all the cases the distribution follows a nice power law over more
than one decade. The observed non-monotonous change in the exponent 
from -1.4 to -2.2 is attributed by Field and coworkers 
to the different inter-vortex distances
attained at the various fields. This may be considered analogous to
the influence of grain friction, shape \cite{Frette:1996} and also
type of base \cite{Altshuler:1999} on the similar exponents
describing sandpile dynamics. The authors also report ``$1/f$" noise
in their experiments, finding power laws for low enough field ramp rates.

\begin{figure}
\centerline{\psfig{figure=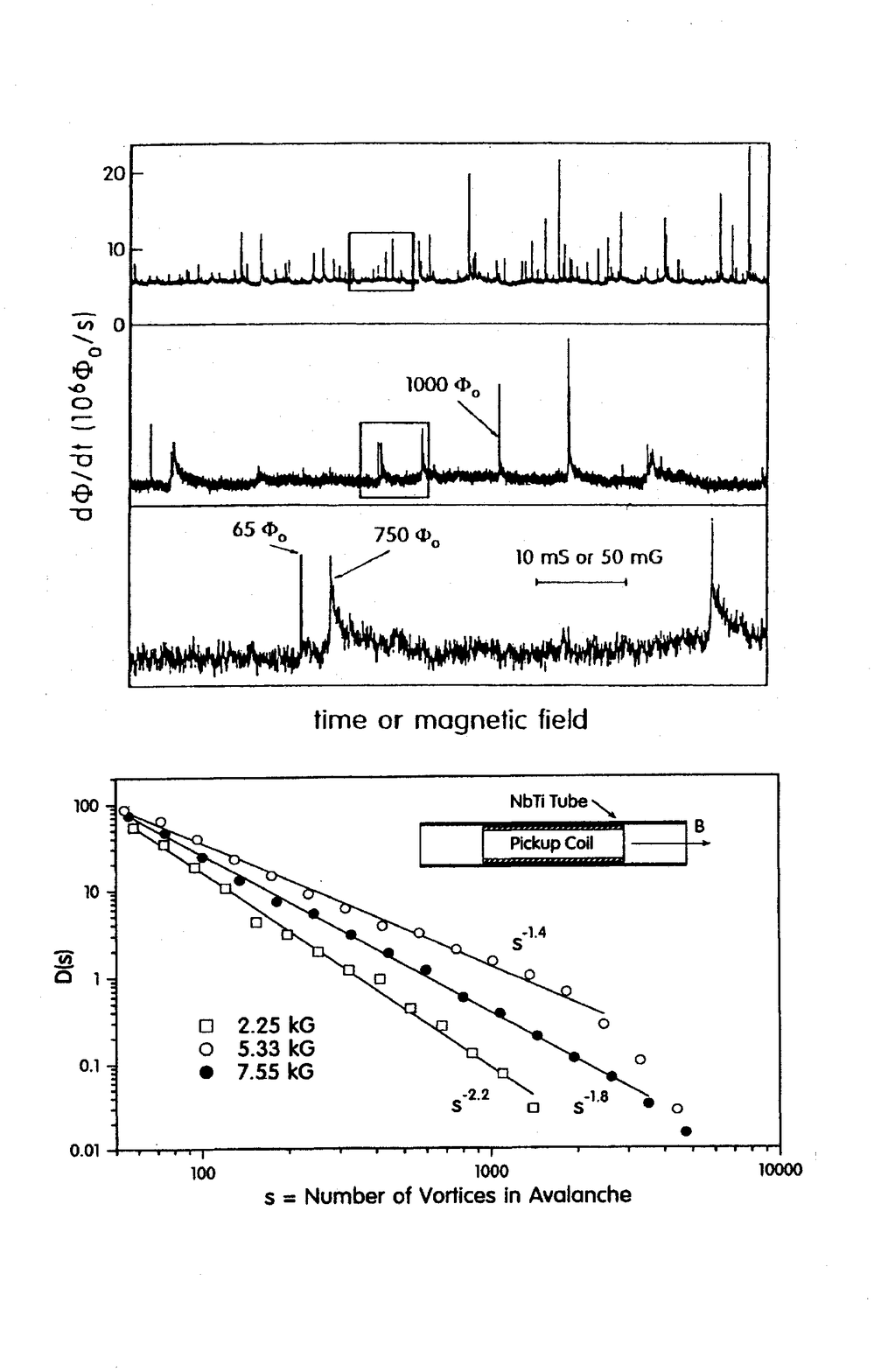,width=8.2cm}}
\caption{Vortex avalanches reported by Field {\it et al.} (1995). 
Upper group of three panels: voltage output for different time windows,
at a field window centered at 7.55~kOe.
Note that the data shown in the small frames in the first and second panels
are shown on an expanded scale in the second and third panels, respectively. 
Lower Panel:
avalanche size distributions for different field windows. 
The inset in this panel shows the experimental arrangement.
\label{f_f4}}
\end{figure}

Let us take a closer look at how the avalanche size was determined
in the work of Field {\it et al.} (1995). Consider a
flux avalanche of length $l$ -- the length along the tube where a
set of vortices ``drops" out of the superconductor and spills into
the hole where the coil is located. Only the corresponding number
of turns, $n =l \, N/L$, where $L$ and $N$ are the coil's total
length and number of turns, respectively, will pick up the flux
change, and the coil responds by inducing the voltage
 $V= n \, ({\rm d} \Phi/{\rm d}t)$,
where $\Phi/\Phi_0$ is the number of vortices participating in the
event. From this the authors defined the avalanche size as an
``effective bundle volume" given by $s \approx l \Phi = (L/N)
\int{V {\rm d}t}$. This is a convenient definition since it was
not possible to determine $l$ directly from the
measurements.\footnote{Detailed experiments in by Heiden and Rochlin (1968) 
already suggested
these kind of limitations in the pick up coil setup.}

We suggest that the avalanche length $l$ 
can be estimated using the collective pinning theory
\cite{Larkin:1973,Larkin:1979}.
According to it, the elastically deformed
vortex lattice is characterized by the length $L_c^b$ and
$R_c$ along, and normal, to the field direction, respectively
\cite{Blatter:1991,Blatter:1993}. Over this volume the vortices
are collectively pinned and behave essentially as one bundle.
The value of $l$ can be evaluated through the simple formula
$l \approx L_c^b \approx (\lambda^2 \xi^3/a_0^4)  (J_{0}/J_{c})^{3/2}$,
where $a_0$ is the inter-vortex distance and $J_0$ is the depairing
current density ($\lambda$, $\xi$ and $J_c$ are defined in section IIA). 
Substituting typical numbers
for a low-$T_c$ alloy at temperatures below 5~K
\cite{Campbell:1972}, with $a_0$ corresponding to a few kOe field,
we get an $L_c^b$ of a few hundred microns \cite{Altshuler:2001},
i.e., much smaller than the length of the pickup coil.
Interestingly, a very early experiment by
Wischmeyer \cite{Wischmeyer:1967}, where two separate coils -- both
similar to the one used by Field {\it et al.} -- were mounted one
after the other on the  inside of a Nb tube, gave two more or less
uncorrelated signals. The two coils were separated by a gap of 2.5~mm,
supporting the above estimate for the size of the ``avalanching objects."

In spite of the limitations inherent in the
method used by Field {\it et al.} (1995), this paper critically fueled
much of the studies of dynamically driven vortex avalanches in
the second half of the 1990s.

\subsection{Micro Hall probe experiments}

In contrast to the pick up coil technique, Hall probes allow one to
directly measure the size of the ``avalanching object" in flux units.
An avalanche event appears here as an abrupt step in the Hall
signal, and the size of the step represents the change in the number
of vortices populating the area under the probe.
Such experiments were first done by
Seidler {\it et al.} (1993), who with a $2 \times 10~ \mu$m$^2$ area Hall
probe detected avalanches in 70~$\mu$m thick, untwinned YBaCuO crystals
during field ramps at 8~Oe/s. The measurements were made below 1~K,
where they found relatively big events and only above a certain field
threshold. Although size distributions
are not presented in this work, the observations suggest that in this
case the avalanches were thermally driven, i.e., they were flux jumps.

Stoddart {\it et al.} (1993) did similar experiments with slightly
smaller Hall probes on 0.2~$\mu$m thick films of Pb, and later also
on Nb films (Stoddart {\it et al.}, 1995). Here, big avalanches were
observed even in the beginning of the field sweep (ramp rate unknown),
but again size distributions were not measured, thus preventing a
comparison with SOC. However, from data obtained using a linear array
of 4 micro-Hall probes, the authors could determine the in-plane
correlations of the avalanche behavior. This analysis identified an
average flux bundle radius of $R_c \sim$ 3.4~$\mu$m for Nb at
$T$ = 4.5~K, in good agreement with the collective pinning
theory.

Zieve {\it et al.} (1996) continued Hall-probe studies of
avalanches in YBaCuO crystals, again performed at very low
temperatures, even well below 1~K. Now the avalanche size
statistics was reported, as well as hysteresis effects observed
when the external field was cycled between 0 and 75~kOe. It was
observed that the steps signaling avalanche behavior have a
distinct onset field, $H_{\rm up}$, during ascent, and that they
disappear on the descending branch at a much lower field. Since
$H_{\rm up}$ is found to be essentially independent of the field ramp
rate, Zieve {\it et al.} (1996) exclude the case 
that the events are thermally driven.
The avalanche size distributions turn out not to have power-laws,
but to be instead sharply peaked around large size (750 vortices)
events, which is indicative of flux jumping, and which is 
definitely not consistent
with the SOC. Nevertheless, Zieve {\it et al.} (1996) argue that their
avalanches are dynamically driven, and that a sandpile analogy serves
to explain the observed hysteretic behavior: It is not equivalent
to add grains to a pile (to increase the field) or to remove
grains from its base (to decrease the field), because the
overall weight of the pile is supported mainly by the grains at
lower positions. To account for the peaked size distributions the
authors extend the analogy. In their opinion, vortex mass
renormalization \cite{Blatter:1993} takes place at the very low
temperatures of these experiments, making vortex inertial effects
significant -- and closer to some sandpile experiments, which show
periodic avalanche events \cite{Held:1990, Rosendahl:1993}.

While SOC behavior was clearly {\it not} found in the experiments
of Zieve {\it et al.}, it is not equally obvious that their
explanation is fully germane: it is today believed that inertial
effects are negligible even at these low temperatures
\cite{Vinokur:2001}. An alternative explanation is provided
by Pl{\'a} {\it et al.} (1996) and others \cite{Olson:1997,Olson:1997a}, 
whose MD simulations suggest that
broad pinning centers with low density --as
expected for the samples measured by
Zieve {\it et al.} (1996)-- produce peaked distributions of avalanches,
while sharp and dense pinning --as
expected for the samples measured by Field {\it et al.}--
produces distributions closer to a power law. 


\begin{figure}
\centerline{\psfig{figure=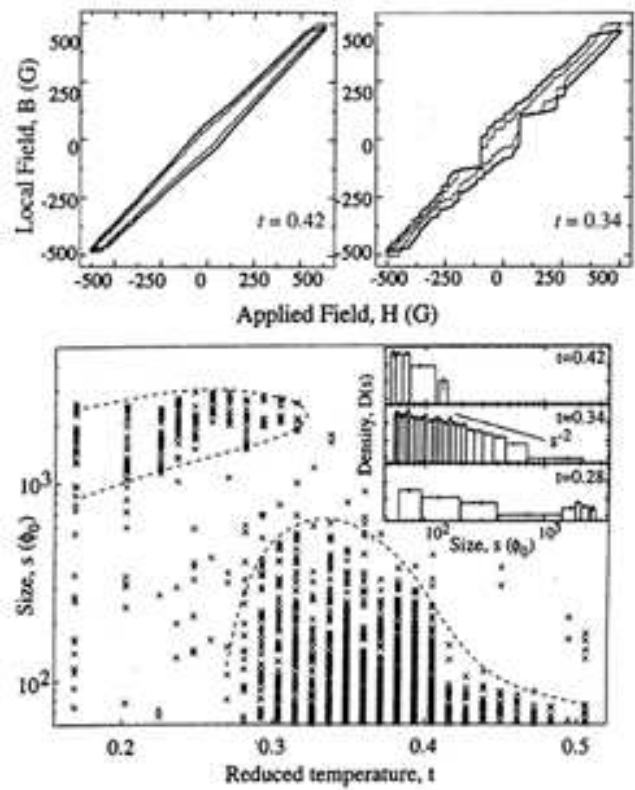,width=8.5cm}}
\caption{Vortex avalanches reported by Nowak {\it et al.} (1997). Upper
panels; local field {\it vs.} applied field for two normalized
temperatures defined as $t=T/{T_c}$ (note the similarity with Fig.~1c).
Lower panel; avalanche size {\it vs.} temperature
diagram, and avalanche size distributions for different
temperatures (inset).
\label{f_f5}}
\end{figure}

Returning to low-$T_c$ materials,  Nowak {\it et al.} (1997)
studied avalanches in Nb films of thickness $d =$~500~nm. Their
samples had an annular shape, with inner and outer diameters of
15~$\mu$m and 0.1~mm, respectively. 
Two 3~$ \times~ 5~ \mu$m$^2$-sized 
Hall probes were used, one mounted over the central hole,
and one at a position 22~$\mu$m off-center, allowing detection of
the {\it total} flux involved in avalanches crossing the inner
edge of the ring (center probe), and the local avalanche activity
in the interior of the sample (off-center probe). Fig.~5
contains the main results of Nowak {\it et al.}, where the upper
two panels show how the local field varies as the applied field is
cycled between $\pm$500~G. The loops, obtained at different
temperatures $t = T/T_c$, both contain distinct steps, and it is
also evident that the magnitude and frequency of these avalanche
events depend strongly on temperature. Moreover, by comparing the
curves from the two probes (thick and thin line represent the
center and internal probe, respectively), one finds them not
always correlated, showing that both global and local flux
avalanches indeed take place. The temperature dependence of this
behavior is compiled in the lower part of the figure, where the
main graph is a scatter plot over all the events detected by the
center probe during two field cycles at each temperature. One sees
that in a narrow range $0.3 < t < 0.4$ the distribution of
avalanche sizes is broad and covers 1-2 decades. At lower
temperatures $0.2 < t < 0.3$ the events cluster at large
system-spanning sizes, typical for thermally triggered
jumps \footnote{This situation was also found in thin $Nb$ films,
although avalanche size statistics were not reported
\cite{Esquinazi:1999}.}, and interestingly one finds at even lower
$t$ that the sizes again become broadly distributed. At $t > 0.4$
only small avalanches occur, and the size distribution is
monotonous and fits a decreasing exponential, as reported also
earlier in \cite{Heiden:1968}. From the figure insets, one sees
that a power law $s^{-2}$ describes the distribution at $t =
0.34$. In this work, also the ramp rate dependence of the
avalanche activity was explored. In their range of rates, from
2 mOe/s to 20 Oe/s, the behavior remained unaffected, showing
that the system is in the slowly driven regime.

Nowak {\it et al.} (1997) explain these data on the basis of
a thermally triggered mechanism. The analysis makes quantitative
use of the stability parameter $\beta$, and both the
superconducting film and the substrate are assumed to absorb heat.
For the particular sample in this
study one has unstable conditions from the lowest
temperatures as well as up to $t=0.37$, which is fully consistent with
the numerous large-$s$ events in this range, as well as the rapid
cut-off of large avalanches at higher $t$.
The broad distribution of avalanches observed in the neighborhood
of $t=0.37$ is related to $\beta$ becoming marginally greater
than 1. Such a fine tuning of parameters may evidently give
power-law behavior, at least over a size range of one decade or so.
An alternative explanation for these findings is given by Olson
{\it et al.} (1997) based on MD simulations. These authors suggest
that, at low temperatures, pinning is so strong that 
interstitial motion of vortices takes place, resulting in 
peaked distributions of avalanche sizes. At higher temperatures
the pinning decreases, so ``pin-to-pin" vortex flow is allowed,
giving rise to wide distributions of avalanche size closer to
a power law.

While the ring configuration of Nowak {\it et al.} appears elegant,
it should be emphasized that the critical state in thin films
placed in a perpendicular applied field deviates quite dramatically
from the picture drawn in Fig.~1. In particular, for a ring-shaped
superconductor, the central hole will contain a sizable
non-uniform field due to shielding currents induced
near the inner edge \cite{Brandt:1997}. Actually, as the applied
field is ramped from zero, there will be two flux fronts
 - one from each edge - advancing into the ring.
The penetration from the inner edge consists of anti-vortices,
because the edge field is here opposite to the applied field.
As the field increases the two fronts eventually meet
(for the Nowak {\it et al.} geometry this occurs at $\sim 3~\mu$m
from the inner edge) and annihilition of the two vortex species
takes place. We find that the actual field when this occurs
is $H_c \simeq J_c d \simeq 150$~G, if we assume a value of 
$J_c = 2 \times 10^6$~A/cm$^2$ for the Nb film. It is clear
that the sample of Nowak {\it et al.} was cycled through a set
of magnetized states with quite complicated flux distributions,
where the purely geometrical (or demagnetization) effects may
prevent drawing direct analogies to sandpile dynamics.

\begin{figure}
\centerline{\psfig{figure=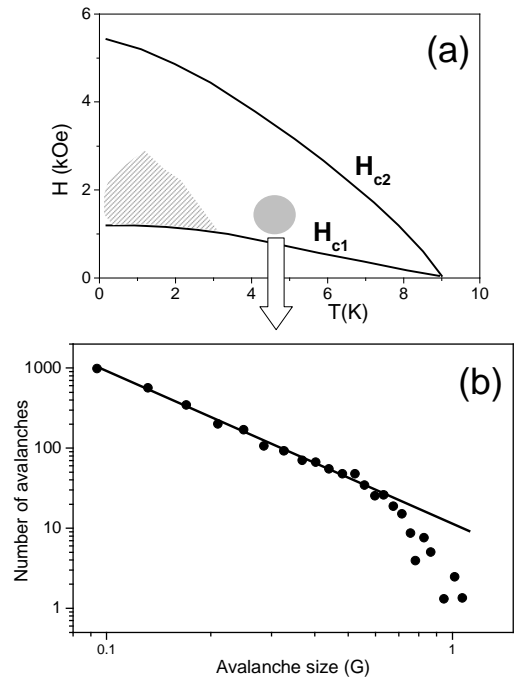,width=7.0cm}}
\caption{Vortex avalanches reported by Behnia {\it et al.} (2000). (a)
Catastrophic avalanches take place in the dashed area of the
$H-T$ diagram, while small ones occur in the rest of the region
between the two lines. (b) Typical avalanche size distribution
corresponding to the small-avalanche region indicated by the gray
circle in the H-T diagram.
\label{f_f6}}
\end{figure}

The first spatial-temporal study of internal vortex avalanches was
made by Behnia {\it et al.} (2000), who made their
measurements on a 20~$\mu$m thick foil of Nb cut as a square with
sides of length 0.8~mm. Unlike previous studies, these
authors even explored the whole $H-T$ region between $H_{c1}$ and
$H_{c2}$ (see Fig.~6(a)). At low temperatures, indicated by the
hatched area, they found catastrophic, flux jump-like avalanches.
Outside this region the behavior was qualitatively different, as
exemplified by the results of the following experiment made at
4.8~K with an applied field around 1.5~kOe (see the circle in the
phase diagram).

A 0.35~mm long Hall probe array consisting of 8 equally spaced
$20 \times 5~\mu$m$^2$ probes, each one with a sensitivity of
0.16~$\Phi _{0}$,  was mounted on the sample along a mid-normal
to one of the sides. After checking that the field creates a
Bean model flux density profile -- something that was difficult to assess
in previous experiments due to the small numbers of Hall sensors --
Behnia and coworkers made a series of measurements as the field was
increased from 1.5~kOe at the rate of 1.1~Oe/s.
>From each probe, they found a local field varying in steps, much
like those reported by Nowak {\it et al.} (1997). The avalanche size
statistics obtained by analyzing the signal from one probe is
shown in Fig.~6(b). In the
small-event region the size distribution follows a power-law
with an exponent of -2.1 (fitted line), which is within the
range of exponent
values reported by Field {\it et al.} (1995). Deviations from the
straight line start around 0.6~G, and reflect a clear deficiency
of large size events. Note that the largest avalanche event
is a field step of 1.1~G, corresponding to a sudden entry of 5
vortices under a probe area already populated by more than 6000
vortices. The authors leave the lack of big avalanches an open
question. Could failure to wait for the extremely rare
events be the simple explanation?

Behnia and coworkers investigated also the temporal correlations
of avalanches by comparing the signal from Hall probes located a
distance 50~$\mu$m from each other. They estimated an average
transit time of 0.8~ms, which gives an avalanche speed of a
few~cm/s. This can be compared with the velocity of vortex motion
during flux flow, given by
 $v \sim \rho_{n} J_{c} / \mu_{0} H_{c2}$, where $\rho_n$
is the normal state resistivity. This gives velocities in the
range of 25-8000~cm/s for parameters near the measuring conditions
of Behnia {\it et al.} (2000). Since thermal activation and a possible
current dependence of the resistivity would decrease this estimate,
we conclude that the velocities of these avalanches,
which have a broad size distribution, are consistent
with a simple picture of the vortex motion, and in strong
contrast to the ultra-fast dendritic flux
penetration discussed later (in Section~C).

Pushing  the Hall probe technique even further,  James
{\it et al.} (2000) used a high resolution SHM to look at
flux penetration into a 1~$\mu$m thick
Nb film shaped as a 100~$\mu$m wide strip. As the applied
field was slowly swept up and then down, they found (by keeping
the sensor stationary 25~$\mu$m from the edge) a step-like behavior
in the Hall signal, much as in previous observations.
But new aspects of the behavior were uncovered when the probe
was scanned across a large part of the sample area. This showed
that the flux does not penetrate with a smooth advancing front,
but instead as a series irregularly shaped protrusions.
These protrusions were easily distinguished from the much larger
and blob-shaped flux patterns that sometimes form abruptly during
field sweeps at temperatures below 4~K. Whereas the blobs are firmly
believed to be the visible result of conventional flux jumps,
James {\it et al.} speculate about the origin of the numerous protrusions,
which are apparent at all temperatures up to $T_c$. A key observation
is that when the protrusions invade the flux free Meissner
region the neighboring ``fingers" show a strong tendency to avoid
each other. Had the protrusions been the fingerprint of scratches or
other defects facilitating easy flux penetration in the film,
this kind of behavior would be very unlikely.
Instead, James {\it et al.} suggest that some long-range repulsive force
plays a role here, and indeed such an interaction does exist between
vortices in thin samples. In contrast to the exponential dependence
in bulk, there is for thin superconductors in a perpendicular field
a long-range inverse distance squared decay 
of the vortex-vortex force due to their surface screening currents  
\cite{Pearl:1964}. Therefore, it may well be that the
flux penetration in the form of these protrusions is an example of
a dynamically driven vortex system full of avalanche dynamics.
By taking differences of the penetration pattern at two fields
differing by 10~G, it was
demonstrated that the flux front advances by an apparently
random sequence of localized bursts of flux motion. The size of these
events was found to vary, but James {\it et al.} (2000) do not report quantitative
size statistics of any kind.

So far, all the mentioned studies of vortex avalanches and their
statistics, i.e., those where SOC ideas were examined using
micro-Hall probes, have lacked knowledge
about the actual ``magnetic landscape" in which the probes were
located. Furthermore, the number of recorded avalanches have been
fairly limited, estimated to be around 150 events in the experiments of 
Zieve {\it et al.} (1996)
and 5000 events in those of Behnia {\it et al.} (2000), 
and  thus hardly sufficient to
convincingly establish power laws when broad size distributions
are found. Both these shortcomings were largely improved by
Altshuler and coworkers \cite{Altshuler:2002} who combined MOI
with the recording of many long series of Hall probe data. Also
the sample used was a Nb foil, 30~$\mu$m thick and cut into a
square with 1.5~mm sides. Figure~7 shows an MOI picture of flux
penetration into the sample, and reveals that the distribution
does not correspond to a simple sample-spanning critical state,
but rather to a set of flux ridges, each having an ``inverted-V,"
Bean's-like profile. In this landscape an 11 probe Hall array,
with $10 \times 10~\mu$m$^2$ sensor areas, was mounted on the
slope of the largest ridge, as indicated by the set of white dots
\footnote{MOI experiments were recently made by the authors
(specifically for this Colloquium) on Nb foils kindly provided by K.
Behnia. It was found that for samples similar to the ones studied
in \cite{Behnia:2000} the flux penetration is globally
non-Bean-like and quite similar to the one seen in Fig.~7, at
least below 500 Oe.} in Fig. 7.

\begin{figure}
\centerline{\psfig{figure=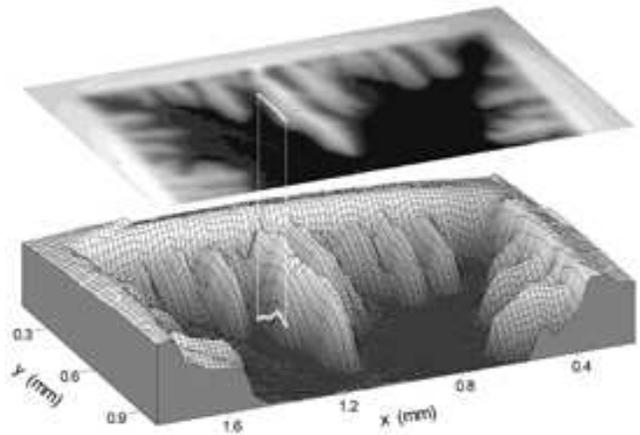,width=8.5cm}}
\caption{Magnetic landscape in a Nb foil where an array of
micro-Hall probes (white dots) detect avalanches coming down the
slope of the largest ``flux ridge" \cite{Altshuler:2002}. \label{f_f7}}
\end{figure}

\begin{figure}
\centerline{\psfig{figure=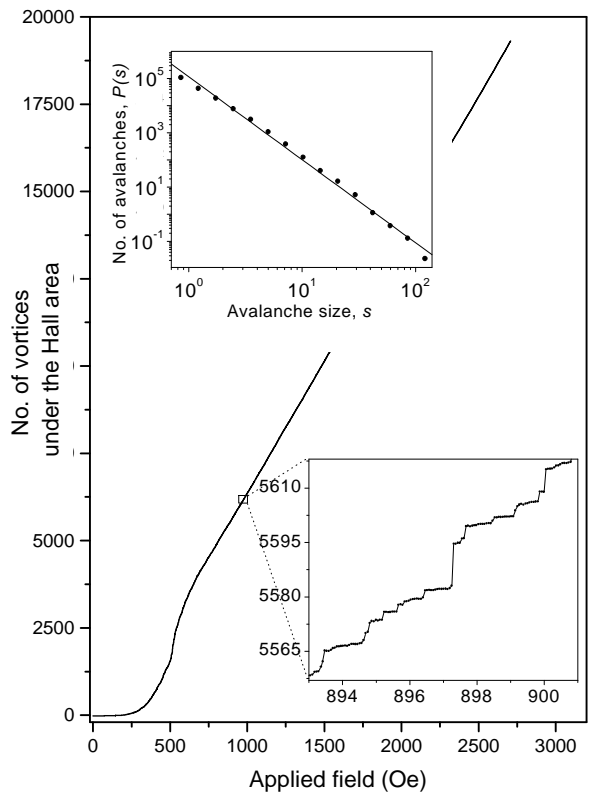,width=8.0cm}}
\caption{Vortex avalanches reported by Altshuler {\it et al.} (2002). Main
curve and lower inset: evolution of the number of vortices under
the Hall probe areas seen in Fig. 7 as the external field is
increased. Upper panel: resulting avalanche size distribution.
\label{f_f8}}
\end{figure}

Shown in Fig.~8 is the signal from one of the Hall sensors
recorded as the field was ramped from 0 to 3.5~kOe at 1~Oe/s and
$T=4.8$~K. When the curve is examined in detail (see lower inset),
one finds clear signatures of avalanche dynamics along the whole
range of fields. By analyzing the data from all of the 11 probes for
repeated numbers of experiments made under the same conditions,
several hundred thousand events were registered and analyzed. The
resulting size distribution is plotted in the upper inset of the
figure, which shows that the avalanche sizes covering two decades
follow a power law with a slope of $-3.0\pm 0.2$. To check the
robustness of this result, the authors explored the avalanche
behavior at many locations by remounting the Hall array at various
positions in the landscape. A power law behavior was found
everywhere and the exponent was essentially the same. The
observed robustness gives grounds for the claim to have, for the
first time, observed SOC in flux dynamics.

In an attempt to investigate also the rigidity of the
vortices involved in these avalanches, a pair of Hall arrays were
mounted on the two sides of the Nb foil with the probes directly
facing each other \cite{Altshuler:2002}. The analysis of
cross-correlations in these data indeed shows {\it some} degree of
correlated behavior on the two sides of the sample, which is most
clearly seen for the bigger avalanches.

Very recently, Radovan and Zieve (2003) used a micro Hall 
probe of $400~\mu$m$^2$ area to look at the avalanche behavior in type II, 
Pb thin films of 100~nm thickness. The external field was slowly ramped up 
to 400~Oe, at various temperatures between 0.27 and 
5.9~K. The authors found large avalanches at relatively high temperatures,
and ``micro-avalanches" at lower temperatures. Based on these observations
they report power law distributions of avalanche sizes at the two temperatures 
0.3~K and 4.3~K,
with exponents of 2.0 and 1.1, respectively.

Another three recent papers report avalanches 
observed by micro Hall probes, although without including avalanche 
size statistics.
Shung {\it et al.} (1998)
found non-catastrophic vortex avalanches on a single
crystal torus made from the heavy fermion superconductor UPt$_3$. 
The authors suggest that the observed sharp 
temperature onset for the
appearance of avalanches is an indication of broken time
reversal symmetry. Ooi {\it et al.} found
signs of SOC in the $1/f$ noise spectrum they obtained from the analysis of 
avalanches found in Bi$_{2}$Sr$_{2}$CaCu$_{2}$O$_{8}$ single
crystals \footnote{These experiments cannot be easily compared 
to others presented in this Colloquium, since they do not
involve a slow increase of the applied field at a fixed temperature.}. 
The same kind of samples were studied also by Milner (2001),
who below 1 K and up to 17 T found huge avalanches that 
strongly resemble those reported by Zieve {\it et al.} for YBCO
crystals. Milner proposes a number of possible 
explanations to the phenomenon, ranging from domain 
structures that modulate the interplay between interpin
and intervortex spacings, to broken time reversal symmetry
in his samples.

\subsection{Magneto-Optical Imaging experiments}

The use of the space- and time-resolving power of MOI to
study flux motion was pioneered already in the
1960s. Inspired by the visualization work of DeSorbo and
Newhouse (1962), Wertheimer and Gilchrist (1967) used a fast camera
technique to study how flux penetrates into disks of Nb, V and various
alloy superconductors. As the applied field increased,
they found events of abrupt flux invasion
starting from a point along the perimeter.
For the understanding of the nature of these avalanches,
one particular observation was crucial,
namely, that the events were accompanied by bubbles formed
in the liquid coolant right above the sample surface.
It was evident that thermo-magnetic flux jumps had, for
the first time, been directly visualized.
These early experiments showed also that the bursts of flux
motion fall into two categories:
``smooth" and ``irregular" (or branching), referring to the
geometrical shape of the invading flux front. The two types
of avalanches were by Wertheimer and Gilchrist (1967) 
found to be related to
the sample quality:
smooth jumps were typical for ``pure" samples, while
the branching patterns were seen only in the alloy disks, suggesting
that material inhomogeneities drastically perturb the
course of the avalanches.

Then in 1993, the branching scenario of flux penetration was
revisited by Leiderer {\it et al.} (1993) making
full use of the high spatial and temporal resolution offered by
the ferrite garnet indicator films. A typical pattern, this time
observed in thin films of YBaCuO, is shown in Fig.~9(a). These
magnificant dendritic patterns were triggered by perturbing a
flux-filled remnant state with
a laser pulse fired at a point near the sample edge. This heated
spot became the root of the branching structure, which is where
the trapped flux has escaped the sample. The study revealed that
if the experiments were repeated in exact detail, the branching
forms would nevertheless vary widely. In other words, these events produce
``irregular" flux patterns that are {\it not} controlled by
quenched disorder in the sample.

Soon after, Duran {\it et al.} (1995) found essentially
the same spectacular behavior in films of Nb. This time the
dendritic flux patterns were produced by simply lowering
the field from 135~Oe applied during the
sample's initial cooling to various temperatures below $T_c$.
These films were 500~nm thick, and the overall conditions resemble
closely the descending field branch in the Hall probe experiments
of Nowak {\it et al.}. What the MOI revealed was that the
dendritic patterns actually vary in their morphology, changing
from quasi-1D structures at temperatures below 0.35~$T_c$, to
highly branched structures {\`a} la the one seen in Fig.~9(a) at
temperatures approaching 0.65~$T_c$. These findings strongly
suggest that the cluster of large-size events at the lowest
temperatures reported by Nowak {\it et al.} (1997) are due to the abrupt
formation of such macroscopic dendritic structures.

\begin{figure}
\centerline{\psfig{figure=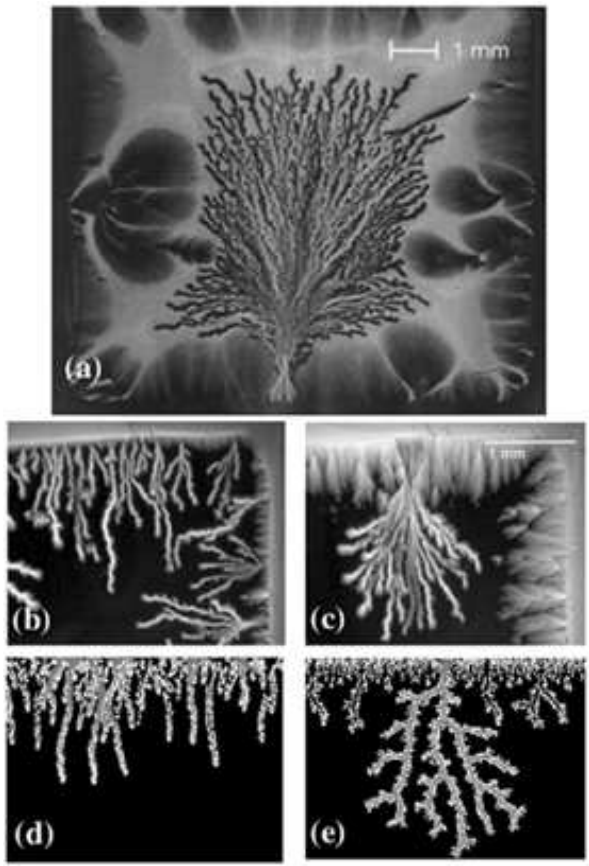,width=7.8cm}}
\caption{Flux dendrites formed abruptly in thin film superconductors.
(a): In YBaCuO at $T = 4.2$~K; (b) and (c): In MgB$_2$ at
$T = 3.8$ and 10~K, respectively; (d) and (e): By vortex dynamics
simulations made for low and high temperature (see text).
\label{f_f8}}
\end{figure}

Dendritic avalanches with the same qualitative
characteristics were observed quite recently also in films of
MgB$_2$  \cite{Johansen:2002}, and Nb$_3$Sn \cite{Rudnev:2003}
only here, as in the very early MOI experiments,
the abrupt events were triggered simply by ramping up
the applied field. During slow ramps after
zero-field-cooling to 4~K, the films became
invaded by numerous dendrites, which 
burst into the Meissner state region one at a time (see Fig.~9(b)).
for the  case of MgB$_2$.
Near 10~K, the dendritic structures become much larger,
as in (c), whereas at even higher temperatures and up
to $T_c$ = 39~K such ``irregular" features cease to be
formed. What is the nature of this type of avalanches, and why
do they take the form of branching flux dendrites? To
find the answer, one should note from Fig.~9 that
the dendrite fingers have a strong tendency to avoid
overlapping. As discussed
in relation to the work of James {\it et al.} (2000), this is
probably a result of the long-range action of the
repulsive force between vortices in thin films. The
same ``explosive" force can possibly also be
responsible for the branching itself, although the mechanism for
selecting these seemingly random bifurcation points
is not yet clear.

These observations formed the basis for a MD-type of computer
code Johansen {\it et al.} (2002), where the dynamical
equation (\ref{MD}) was modified to account for the thin film
geometry, i.e., by using $1/r^2$ intervortex forces, and
adding a term for the Lorentz force from the Meissner currents,
which in thin superconductors flow over the whole area. Finally, a
thermal component was introduced: When any vortex $i$
moves a distance $\Delta r_i$, given by evaluating $v_i$,
an amount of heat, $Q_i=\Delta r_i f_i$, is produced that
raises the temperature in the neighborhood of the trajectory by
 $\Delta T \propto Q_i$. This has then a direct effect on
the local pinning conditions, since the pinning force is taken
to be $T$-dependent (as $f_i^{vp} \propto 1 - T/T_c$). Results of these
simulations are seen in Figs.~9(d) and (e), showing flux
penetration patterns corresponding to low and high temperatures,
respectively. Notice that some of the dendritic fingers have a
``spine," which is the instantaneous map of the temperature rise
due to recent traffic of vortices penetrating from the upper edge.
Evidently, the avalanche morphology found experimentally is very
well reproduced by these simulations. Also analytical efforts have
addressed the same question, and calculations by Aranson {\it et
al.} (2001) suggest that vortex ``micro-avalanches" can be triggered by a
hot spot, and that the temperature distribution can evolve in a
branching manner. Despite the qualitative
success of the theoretical work, more needs to be done to
understand these avalanches at a quantitative level. For example, 
MOI using double-pulse laser illumination with
time intervals less than 10~ns has shown that the speed of
dendrite propagation in YBaCuO is close to 25~km/s \cite{Bolz:thesis}. 
This is orders of magnitude higher than the avalanche velocity reported
by Behnia {\it et al.} (2000), and actually the two scenarios appear 
totally different, as one would expect for dynamically and thermally 
driven systems. Interestingly, the speed of dendrite propagation even
exceeds the sound velocity in the material, raising questions about 
which non-phonon heat conduction mechanism is here active.

Very recently, MOI was used to study also non-catastrophic
avalanches. In the work of Bobyl {\it et al.} (2003) the first spatially 
resolved observation of vortex avalanches on a mesoscopic scale is reported. 
A thin film of MgB$_2$ was investigated at temperatures below 10~K, where 
flux dendrites can form in this material, but the applied field was 
now kept below the 
threshold for dendrite formation. By increasing the field slowly
(60~mOe/s) avalanches were observed by subtracting subsequent 
images recorded at intervals of $\Delta H = 0.1$~Oe. 
All the avalanches were seen to have a regular shape with no sign of 
ramification, and they appear at seemingly random 
places mainly near the edge of the film.
The total number of vortices participating in an 
avalanche varied between 50 and 10000. 
However, the work does not report any detailed statistics.
Interestingly, the mesoscopic avalanches, having a typical linear 
size of 10-20 $\mu$m, continue to form also at 
the higher fields where the large dendrites dominate the 
flux penetration. Moreover, it is found that, above 10~K, 
both types of avalanches
(mesoscopic ones and dendrites) cease to form suggesting that 
only one physical mechanism is responsible for both.

In a work by Aegerter {\it et al.} (2003) an 80~nm thick film 
of YBaCuO was, after zero-field cooling to 4.2~K, subjected to a 
perpendicular field slowly increased in a stepwise manner. After each 
field step of 0.5~Oe, the sample was allowed to relax for 
10~seconds before an image was taken.
By subtracting subsequent images, the difference in flux density 
$\Delta B_z(x,y)$ was obtained and integrated over a sub-area 
$L \times L$ of the total 
field of view. This revealed clearly that the evolution of the magnetic
flux inside the sample is intermittent with occasional bursts of 
various sizes. To allow for a finite-size scaling type of analysis, 
the authors let $L$ vary between $180~\mu$m and $15~\mu$m.
The histogram of avalanche size distributions with 4 different $L$ values
shows power laws, which when combined extend over more
than 3 decades. Furthermore, plotting the histogram versus the scaled 
avalanche size $s/L^D$ shows a good data collapse 
using $\alpha = 1.29$ and $D = 1.89$. In addition, the authors measure both 
the so-called roughness exponent and the fractal dimension of the avalanche 
clusters, and show that the set of exponents obey a universal scaling relation.
This gives strong indication that SOC is present in their system.

Related to this is the earlier observation 
of kinetic ``roughening"  of advancing flux fronts
in high-$T_c$ films (Surdeanu {\it et al.}, 1999).
By applying scaling analysis, it was shown that there exists two regimes;
at small length scales or short time scales, where static disorder
dominates, the roughening and growth exponents correspond to a
directed-percolation-depinning model, whereas at larger scales
temporal stochastic noise dominates and the exponents come close
to those of the Kardar-Parisi-Zhang (KPZ) model. This finding has
common ground with finding of
the dynamically driven avalanche community: theoretical models in
sandpiles have established relations between the critical
exponents of avalanche dynamics and those for interface growth,
including for the KPZ \cite{Paczuski:2000,Chen:2002}.

The MOI technique made a giant leap forward
when Goa {\it et al.} (2002) succeeded to resolve individual
vortices, and thereby directly visualizing their motion.
Immediately, one obtained here a new method capable of following
vortex avalanche dynamics in full detail,\footnote{Compared
to Lorentz microscopy, the only other method with the same
capability, the MOI is not restricted to samples so thin
that the electron beam goes through.} and not only through
sampling of the flux density integrated over some area.
In particular, single-vortex resolution MOI could contribute
to test experimentally the role of interstitial versus ``pin-to-pin"
motion of vortices during avalanches, as predicted in MD simulations
\cite{Olson:1997,Olson:1997a}, or even the details of 
``braided rivers" of vortices resulting from CA simulations
\cite{Bassler:1999}.To illustrate what is now 
possible, Fig.~10 shows an image
which is the difference of two MOI pictures recorded before
and after the applied field was increased by 4~mOe during
1~second. The superconductor is here a 0.1~mm thick single
crystal of NbSe$_2$ at 4~K. The bright and
dark dots show the local increase and decrease of the field,
i.e., they are the positions the vortices have hopped to and
from, respectively. The areas where such dots are absent also
contain vortices, but they have not moved during this
particular interval. From the image one can clearly
identify vortex avalanches of various sizes, e.g., there is
a quite large event taking place on the left side, and many
small ones, down to individual hops, are scattered over the
whole field of view.
Although this new high-resolution MOI method has not yet been
used specifically to study avalanche aspects, it is
evident that the experimental potential is huge,
and will bring us closer to a full understanding of
vortex dynamics.

\begin{figure}
\centerline{\psfig{figure=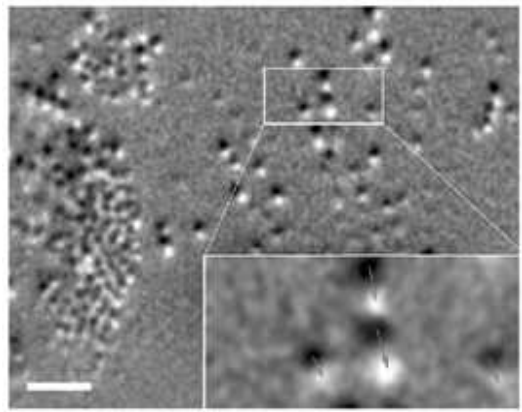,width=7.0cm}}
\caption{Vortex avalanches in NbSe$_2$ observed by MOI.
The bright and dark dots show where vortices have moved to
and from, respectively, during a field step of 4~mOe.
The scale bar is 10 micron long.
\label{f_f9}}
\end{figure}

\subsection{Miscellaneous experiments}

Vortex avalanches not associated with conventional 
flux jumps have been detected also
through other techniques
like SQUID Magnetometry \cite{Aegerter:1998,Kopelevich:1998,Wang:1993}
and torque magnetometry \cite{Hope:1999}. Among them, only 
Aegerter (1998) reported avalanche size distributions
by studying the detailed
flux motion during creep in a Bi$_{2}$Sr$_{2}$CaCu$_{2}$O$_{8}$ crystal. 
Instead of
driving the vortices 
by increasing the 
applied magnetic field, the avalanches were
here created by thermal activation in a constant applied field.
Using a SQUID sensor the events were recorded as time went by for
more than $10^5$ seconds. The main finding is that at low
temperatures (0.06~$T_c$) the distribution of avalanche sizes
shows power law behavior, whereas at higher temperatures (0.8~$T_c$) the
distribution becomes exponential. A theoretical discussion of
these results has been provided by Mulet {\it et al.} (2001), based on the
Bassler-Paczuski CA with additional Monte-Carlo rules to account
for the slow thermal activation. They conclude that the critical
exponents obtained in creep experiments can be related to, but are
not identical to, those predicted in the original SOC scheme.

To summarize, Table~1 gives an overview of
the main results concerning avalanche statistics 
and experimental conditions reported in all the
papers reviewed in this Colloquium.

\begin{widetext}

\begin{table}
\begin{center}

\caption{List of experiments reporting vortex avalanche size
distributions. The information in the table was extracted directly
or indirectly from the references cited.
``Exp," ``peak," ``power"(exponent) and ``stexp" refer to exponential,
peaked, power-law and stretched exponential distributions of avalanche
sizes, respectively.}

\begin{tabular}{ccccccccc}

\hline\hline $Ref$ & $Geom.$ & $Material$ & $Sensor$ &
$Avalanche$ & $T/T_c$ & $H\ range$ & $Rate$ &$Avalanche$ \\
&  &  &  & $type$ &  & [kOe] & [Oe/s] & $distribution$\\ \hline
Heiden {\it et al.} & hollow & Pb-In & pickup & off-edge
& 0.6 & 0.55 -- & 10 -- 100 & exp \\
(1968) & cylinder &  & coil &  &  & 0.85 &  &  \\ \hline
Field {\it et al.} & hollow & Nb-Ti & pickup & off-edge & 0.3
& 2.25 -- & 5 & power(1.4 -- 2.2) \\
(1995) & cylinder &  & coil &  &  & 7.55 &  & (slow ramps) \\ \hline
Zieve {\it et al.}& planar & YBCuO & Hall & internal &
 $\leq$0.01 & 0 -- 80 & 7 & peak \\
(1996) &  & crystal & probe &  &  &  &  &  \\ \hline
Nowak {\it et al.} & planar & Nb & Hall & off-edge & 0.15 -- &
$-0.5$ -- & 0.002 --  & peak/power(2.0)\\
(1997) & ring & film & probes & \& internal & 1.12 & 0.5 & 20&\\ \hline
Aegerter & planar & BSCCO & SQUID & off-edge & 0.06 -- & ? &
0 & exp/power(2)\\
(1998) & & crystal &  & & 0.8 &  &    \\ \hline
Behnia {\it et al.} & planar & Nb & Hall & internal & 0.52
& 1.5 & $\sim~1$ & peak/power(2.05)\\
(2000) &  & film & probes &  & & & & /stexp\\ \hline
Altshuler {\it et al.} & planar & Nb & Hall
probes & internal & 0.5 & 0 -- 3.5& $\sim~1$ & power(3.0)\\
(2002) & & foil & \& MOI & & & & \\ \hline
Aegerter {\it et al.} & planar & YBCO & MOI & internal & 0.05
& 0-0.15 & $\leq$0.05 & power(1.30)\\
(2003) &  & film &  &  & & & & \\ \hline
Radovan, Zieve & planar & Pb & Hall & internal & $\leq$0.7
& 0-0.04 & 0.2-3.3 & peak\\
(2003) &  & film & probes &  & & & & /power(1.1,2.0) \\\hline\hline

\end{tabular}
\end{center}
\end{table}
\end{widetext}

\section{SUMMARY AND OPEN QUESTIONS}

When an account of a given scientific subfield is written long 
after the key developements, a mysterious filtering process takes place
which results in a nice {\it concerto} of experiments, perfectly
aimed at the ``big question." However, in the case of vortex avalanche
experiments cruel reality has forced us to replace such an idyllic approach
by a much more disjointed literary style. 
Nevertheless, we
have still been able to distill from the available experiments a
set of issues and questions that
may contribute significantly to the
understanding of the physics beyond vortex avalanches.

Although low-$T_c$ materials dominate most of the experiments
in which avalanche size statistics are reported, the types of
samples used are widely
different (cylinders, films, foils) and the temperature, field ranges,
and field sweep rates vary quite a bit from report to report.
The occurrence of avalanches in the different regions of the $H-T$ 
phase diagram has been only rarely explored.
In practice, it has proved difficult to tell if the observed avalanches
are thermally or dynamically triggered, although there is
consensus that the first ones abound at $T$ below 4~K or
so --at least in low-$T_c$ samples. Remarkably,
only one experimental work on high-$T_c$ materials reports 
to have found non-catastrophic avalanches during
slowly increasing field. Is this general situation
due to lack of instrumental resolution, or perhaps are the
avalanches just ``smoothed out" by thermal activation?

Even when non-catastrophic avalanches are detected as the field
is swept, opinions are
divided as to their origin. Some authors claim
that Self Organized Criticality is at the core of the dynamics.
But robust,
well-defined power laws have proved somewhat elusive:
Nature does not seem to like more than two decades of 
avalanche sizes measured in a single experiment...or
have we failed to be patient enough to collect the
appropriate wealth of data  \cite{Avnir:1998}?

Some simulations suggest that the type of avalanche size
distribution may depend on the nature and density of pinning sites
 --in analogy with experiments in sandpiles with
different types of grains and bases on which the piles 
are grown. Definitive experiments to
check this hypothesis can be performed only on samples with
artificially tailored pinning landscapes. If true, could measurements of 
avalanche size
distributions become a tool
to figure out the pinning features of a given sample?

In the case of non-catastrophic events, and when power law behavior is found,
there is a great dispersion in the critical exponent of the 
avalanche size distributions. This applies to both experiment and theory: 
While for the first category the exponent ranges from 1.3 to 3.0, 
in the second it typically spans from 1 to 2, and it can go even further.
An important principle question then arises: 
Is it possible to establish a one-to-one correspondence between the
different experiments and models?

Power law distributions of avalanche sizes are expectedly 
associated to linear flux profiles (like originally proposed
by Bean), since nonlinear ones, in principle, cannot
result in scale-invariant avalanches. 
Many of the recent experiments have been made on thin superconductors in a 
perpendicular magnetic field where
the flux density profiles have an enhanced slope near the sample's edge
and center. This applies even for samples with a
constant critical current density. In bulk samples there is also a possibility for
having non-linear profiles due to a $B$-dependence of the
critical current density, e.g. as in the Kim model.
What exactly are the differences in avalanche behaviors when  non-Bean flux 
profiles are present? Are they diminished when the sensors
cover only a small area of the sample?

The very nature of the ``avalanching objects" is sometimes in question
due to the lack of appropriate instruments: are they
individual vortices, or flux bundles?
Are they rigid entities? Or perhaps we are seeing the
irregular growth of tiny
flux fingers, only visible with the most sophisticated instruments?

Imaging techniques suggest that the scenario where avalanches take
place can eventually be quite distant from the basic
Bean's critical state. Catastrophic
avalanches seem to be associated with ``bursting," non-repeatible dendritic
structures, while non-catastrophic ones are mostly found in materials
where the field penetrates as fingers with a Bean's-like
cross section. Even ``roughness" in the critical
state can be related to vortex avalanches,
but this relation is just starting to be properly established. MOI
seems to have the potential to materialize our wildest
dreams in vortex avalanche studies: high spatial
and temporal resolutions, and
the ability to take ``magnetic pictures" of an ample region of the sample.
This technique is only limited by the speed of data
acquisition and data storage
capabilities...but, with a little patience, these
will find their way from Hollywood special effects
departments to scientific labs.

All in all, it becomes clear that there are more questions
than answers in the field of vortex avalanches. This is of course
good news for the scientists working in Complex Systems, but
probably even better news for the vortex physics community, which
is busy these days tightening up the last bolts to the {\it equilibrium}
$H-T$ diagram of superconductors.

\section*{Acknowledgments}

The authors acknowledge useful discussions with Ch. Aegerter, P.
Bak, K. E. Bassler, A. J. Batista-Leyva, K. Behnia, E. H. Brandt, J.
R. Clem, T. Giamarchi, A. Gurevich, H. Herrmann, H. Jaeger, H. J.
Jensen, P. Leiderer, X. S. Ling, J. Luzuriaga, M. C. Marchetti, M.
Marchevsky, R. Mulet, E. Nowak, M. Paczuski, G. Parisi, H.
Pastoriza, O. Ramos, G. Reiter, B.U. Runge, G. Seidler, D. V.
Shantsev, O. Sotolongo, V. Vinokur, R. J. Wijngaarden, Y. Yeshurun
and E. Zeldov. We thank \AA. F. Olsen for MOI work made
specifically for this paper. E. A. thanks A. Rivera, J. Altshuler
and M. {\'A}lvarez for inspiration and support, and also the ACLS/SSRC
Working Group on Cuba for online access to bibliographic materials. 
T. H. J. is grateful to his patient family, and for financial support from
the Norwegian Research Council.

\bibliographystyle{apsrmp}

\end{document}